\documentclass[letterpaper, 10pt, conference]{ieeeconf}

\pdfoutput=1
\IEEEoverridecommandlockouts                              
\usepackage[utf8]{inputenc}
\usepackage[pdftex]{graphicx}
\graphicspath{{./Figures}}
\usepackage{float}
\usepackage{subcaption} 
\usepackage{epstopdf} 
\usepackage{amssymb}
\usepackage{amsfonts}

\usepackage{amsthm} 
\usepackage{cite}
\usepackage{flushend}

\usepackage{threeparttable}
\usepackage{tablefootnote}
\usepackage{multirow}
\usepackage{booktabs}

\makeatletter
\let\NAT@parse\undefined
\makeatother
\usepackage{xcolor} \usepackage{hyperref}
\hypersetup{%
    colorlinks=true,
    linkcolor={blue!50!black},
    citecolor={blue!50!black},
    urlcolor={blue!50!black}
}

\usepackage{amsmath}

\newtheorem{remark}{Remark}

\newcommand{\R}{\mathbb{R}} 
\newcommand{\N}{\mathbb{N}} 
\newcommand{\Ni}[2]{\N_{[#1, #2]}} 
\newcommand{\cc}[1]{{\mathcal{#1}}} 

\def\Sum#1#2{\sum\limits_{#1}^{#2}} 

\def\traj{\cc{T}} 
\def\spaceT{\mathbb{T}} 
\def\tX{\cc{X}} 
\def\tU{\cc{U}} 
\def\tY{\cc{Y}} 
\def\nx{{n_x}} 
\def\nu{{n_u}} 
\def\ny{{n_y}} 
\def\nth{{n_\theta}} 
\def\Tds{\cc{D}} 
\def\nD{{n_p}} 
\def\prefFunc{\Pi} 
\def\surr{\pi} 
\def\model{\sigma} 
\def\settime{{\kappa_{\varepsilon}}}

\def\fracg#1#2{{\displaystyle{\frac{#1}{#2}}}} 

\begin{document}
\title{\LARGE \bf Learning the MPC objective function from human preferences}

\author{Pablo Krupa, Hasna El Hasnaouy, Mario Zanon, Alberto Bemporad%
\thanks{This work has received support from the European Research Council (ERC), Advanced Research Grant COMPACT (Grant Agreement No. 101141351).
The authors are with the IMT School for Advanced Studies, Lucca, Italy. Emails: {\tt\small \{pablo.krupa, hasna.elhasnaouy, mario.zanon, alberto.bemporad\}@imtlucca.it}. Corresponding author: Pablo Krupa.}%
}

\maketitle
\thispagestyle{plain}
\pagestyle{plain}

\begin{abstract}
In Model Predictive Control (MPC), the objective function plays a central role in determining the closed-loop behavior of the system, and must therefore be designed to achieve the desired closed-loop performance.
However, in real-world scenarios, its design is often challenging, as it requires balancing complex trade-offs and accurately capturing a performance criterion that may not be easily quantifiable in terms of an objective function.
This paper explores preference-based learning as a data-driven approach to constructing an objective function from human preferences over trajectory pairs.
We formulate the learning problem as a machine learning classification task to learn a surrogate model that estimates the likelihood of a trajectory being preferred over another.
The approach provides a surrogate model that can directly be used as an MPC objective function.
Numerical results show that we can learn objective functions that provide closed-loop trajectories that align with the expressed human~preferences.
\end{abstract}

\begin{keywords}
    Model predictive control, Calibration, Preference-based learning, Machine learning, Classification.
\end{keywords}

\section{Introduction} \label{sec:intro}

The main advantage of Model Predictive Control (MPC) is its ability to handle constraints while optimizing a given objective function~\cite{Rawlings_MPC_2017, Camacho_S_2013}.
The parameters defining the MPC controller and its objective function need to be tuned to achieve the desired system behavior and performance.
However, tuning MPC parameters remains a challenging task in practice, as it requires expert knowledge and significant manual effort to achieve the desired closed-loop behavior~\cite{Garriga2010ModelPC}.
Moreover, a limiting factor in many industrial applications is that obtaining an explicit measure quantifying the desired performance may be difficult or impossible.
Instead, only qualitative assessments by a human expert may be available.

The challenge of calibrating the parameters of an MPC controller, otherwise known as \emph{MPC tuning}, has received considerable attention from the research community.
Many approaches leverage the availability of an explicit characterization of the desired system performance.
For instance, Bayesian optimization~\cite{Brochu2010} and other global optimization techniques~\cite{Bemporad2020} have been applied to MPC tuning~\cite{Forgione2020},~\cite{Lucchini2020}, or to adapt the learning procedure of the predictive model utilized in MPC to improve closed-loop performance~\cite{Piga2019}.
These techniques, among others such as sensitivity-based approaches, have also been extended to tuning PID and LQR controllers~\cite{Fiducioso2019},~\cite{Marco2016, masti2021tuning}.
Other approaches that have been used for MPC tuning include the use of reinforcement learning~\cite{Gros_ENMPC_RL_TAC_2020, Mehndiratta_tuneMPC_RL_ICIRS_2018} or inverse optimality~\cite{Zanon_IO_TAC_2022}.

Preference-based learning provides a promising alternative when there is no available explicit function characterizing system performance.
Instead, human preferences are used, which can be expressed in different ways: pairwise comparisons, where a human selects the preferred option between two choices; ordered preferences, where multiple options are ranked in sequence; or binary yes/no choices.
A common application that considers this paradigm is training large language models to align with user preferences~\cite{Ziegler2019FineTuningLM, Stiennon2020}.
Different approaches have been employed to learn from human preferences, such as reinforcement learning~\cite{wirth_JMLR_2017, christiano2017deep}, or preferential Bayesian optimization along with Gaussian processes for classification~\cite{Gonzalez2017}.
The result is a surrogate model, typically in the form of a deep Neural Network (NN), that has been trained to capture the underlying preference function driving the human decision-making.

Several preference-based learning approaches have been explored for tuning controllers.
In~\cite{Coutinho2024}, preferential Bayesian optimization is applied to tune proportional-integral controllers from human-in-the-loop feedback providing pairwise preferences.
In \cite{Zhu_ECC_2021, Zhu_TCST_2022}, a semi-automated calibration approach was proposed for MPC tuning based on pairwise preferences between experiment outcomes.
A surrogate model of the underlying preference function is fitted using the global optimization algorithm GLISp~\cite{Bemporad2020GlobalOB} by iteratively proposing new calibration parameters in an exploration-exploitation framework.
Inverse reinforcement learning is another approach that is strongly related to preference-based learning and that has also been used for MPC tuning~\cite{Tao20247294}.

This paper proposes an MPC tuning approach where a dataset of pairwise comparisons of input and state trajectories are used to train a surrogate model that captures the underlying human preference function.
In contrast with~\cite{Zhu_ECC_2021, Zhu_TCST_2022}, we pose the learning problem as a binary classification problem that is solved by standard machine learning tools.
The problem is posed such that the fitted model can be directly used as the objective function of an MPC controller to achieve a closed-loop behavior that aligns with the collected human preference.

The rest of the paper is organized as follows.
Section~\ref{sec:pref} presents the problem formulation, the proposed approach for training a surrogate model of the underlying preference function, and how it is then used as the objective function of an MPC controller.
Section~\ref{sec:results} shows numerical results for a system of oscillating masses, where we fit quadratic objective functions to achieve a closed-loop behavior that aligns with the human preferences.
We conclude with Section~\ref{sec:conclusions}.

\newpage
\noindent\textbf{Notation:}
Given $x, y \in \R^n$, $x \leq (\geq) \; y$ denotes componentwise inequalities.
The exponential function is denoted by $\exp(\cdot)$.
The set of natural numbers (including $0$) is denoted by $\N$.
For $i, j \in \N$ with $i \leq j$, $\Ni{i}{j} \doteq \{i, i+1, \dots, j\}$.

\section{Learning MPC cost from human preferences} \label{sec:pref}

This section presents the proposed preference-based learning process to obtain an objective function for MPC that reflects human preferences.
The idea is to use a dataset containing pairwise comparisons of system trajectories.
The dataset can be used to learn the objective function of an MPC controller such that the closed-loop behavior of the system is likely to be preferred by the human over any other~alternative.

\subsection{Problem formulation} \label{sec:pref:setup}

Consider a system modeled by the discrete-time dynamics%
\begin{subequations} \label{eq:sys}
\begin{align}
    x(t+1) &= f(x(t), u(t)), \label{eq:sys:x} \\
    y(t) &= g(x(t)), \label{eq:sys:y}
\end{align}    
\end{subequations}
where $x(t) \in \R^\nx$, $u(t) \in \R^\nu$ and $y(t) \in \R^\ny$ are the system state, input and output at the discrete-time instant $t \in \N$, and $f\colon \R^\nx \times \R^\nu \to \R^\nx$, $g\colon \R^\nx \to \R^\ny$ define the system state dynamics and output equation, respectively.
Without loss of generality, the control objective is to steer the system state to the origin $x_r = 0$, $u_r = 0$, which is assumed to be an equilibrium of~\eqref{eq:sys}; that is, $x(t) \to x_r$ and $u(t) \to u_r$ as $t \to +\infty$.
Moreover, we assume that the system must satisfy (possibly coupled) state-input constraints $(x(t), u(t)) \in \cc{Z}, \forall t$, for some given $\cc{Z} \subseteq \R^\nx \times \R^\nu$.

Let $\traj = (\tX, \tU, \tY) \in \spaceT$ be a finite trajectory of~\eqref{eq:sys}, i.e., $\tX = (x_0, x_1, \dots, x_N)$, $\tU = (u_0, u_1, \dots, u_{N-1})$ and $\tY = (y_0, y_1, \dots, y_{N-1})$, where $x_i \in \R^\nx$, $u_i \in \R^\nu$ and $y_i \in \R^\ny$ satisfy $x_{i+1} = f(x_i, u_i)$, $y_i = g(x_i)$, $\forall i$, and $N \in \N$.\footnote{We omit $N$ from the notation of $\traj$ for convenience. Throughout the article, all trajectories are of length determined by $N$ unless stated~otherwise.}

We assume that the human expresses their assessments according to a preference function $\prefFunc\colon \spaceT \times \spaceT \to \{0, 1\}$ that encodes a preference between two trajectories $\traj_i$ and $\traj_j$ as:
\begin{equation*}
\prefFunc(\traj_i, \traj_j) =
\begin{cases}
    1\; \text{if}\; \traj_i \; \text{is preferred over} \; \traj_j, \\
    0\; \text{otherwise}.
\end{cases}
\end{equation*}
In our framework, $\Pi$ is unknown but accessible, in that we can evaluate $\prefFunc(\traj_i, \traj_j)$ for any pair of trajectories $\traj_i$, $\traj_j$.
In a real scenario, $\Pi$ could be defined as the preference that an expert operator assigns by looking at two different trajectories.
The exact reason for the expert operator to choose one trajectory over another may be difficult to formally represent, but we assume that there is some underlying $\Pi$ that the operator uses, consciously or unconsciously.

Let $\{\traj_i\}_{i=1}^{n_t}$ be a collection of $n_t$ trajectories $\traj_i \in \spaceT$ of system~\eqref{eq:sys}.
We assume the availability of a dataset
\begin{equation*}
    \Tds = \{(\traj_{\cc{I}_\ell}, \traj_{\cc{J}_\ell}, p_\ell)\}_{\ell=1}^{\nD}  
\end{equation*}
that collects $\nD$ trajectory pairs $(\traj_{\cc{I}_\ell}, \traj_{\cc{J}_\ell})$, where $\cc{I}_\ell$ and $\cc{J}_\ell$ are the indices of the trajectories from $\{\traj_i\}_{i=1}^{n_t}$ considered in the $\ell$-th preference $p_\ell \doteq \Pi(\traj_{\cc{I}_\ell}, \traj_{\cc{J}_\ell})$.

\subsection{Training a surrogate model of the preference function} \label{sec:pref:model}

Our objective is to learn a surrogate model $\surr\colon \spaceT \times \spaceT \to \{0, 1\}$ of $\prefFunc$ that would ideally satisfy
\begin{equation} \label{eq:surr:objective}
    \surr(\traj_i, \traj_j; \theta) = \prefFunc(\traj_i, \traj_j),\; \forall \traj_i, \traj_j \in \spaceT,
\end{equation}
where $\theta \in \R^\nth$ is the vector of parameters defining $\surr$.
Vector $\theta$ will be trained using the dataset $\Tds$ so as to maximize the satisfaction of~\eqref{eq:surr:objective}.
Furthermore, our objective is to be able to use $\surr$ to define the objective function of an MPC controller so as to obtain closed-loop trajectories that would generally be preferred over others according to $\prefFunc$.
To this end, we take
\begin{equation} \label{eq:surrogate}
\surr(\traj_i, \traj_j; \theta) =
\begin{cases}
    1\; \text{if}\; \model(\traj_i; \theta) \leq \model(\traj_j; \theta), \\
    0\; \text{if}\; \model(\traj_i; \theta) > \model(\traj_j; \theta),
\end{cases}
\end{equation}
where $\model\colon \spaceT \to \R$ is parameterized by $\theta$.
We note that this choice of $\surr$ is taken to allow us to use $\model$ as the objective function of an MPC controller, as explained in Section~\ref{sec:pref:mpc}.

The objective of obtaining a function $\surr$ that always satisfies~\eqref{eq:surr:objective} is generally unattainable due to the lack of knowledge of the underlying (and unknown) preference function $\prefFunc$.
However, we can formulate a learning problem of function $\pi$ to maximize the accuracy in satisfying~\eqref{eq:surr:objective} for the trajectory pairs in the dataset $\Tds$.
In particular, we train $\theta$ by solving a binary classification problem using the dataset $\Tds$~\cite{Dreiseitl_log_reg_2002}.
As typically done in logistic regression for binary classification, we use the sigmoid function to model the probability of the preference between two trajectories $\traj_i, \traj_j \in \spaceT$.
We take this probability as
\begin{equation} \label{eq:sigmoid}
    P_\surr(\traj_i, \traj_j; \theta) = \fracg{1}{1 + \exp(\model(\traj_i; \theta) - \model(\traj_j; \theta))}.
\end{equation}
Taking this expression for $P_\surr$, the surrogate model~\eqref{eq:surrogate} can then equivalently be expressed as
\begin{equation} \label{eq:surrogate:v2}
\surr(\traj_i, \traj_j; \theta) =
\begin{cases}
    1\; \text{if}\; P_\surr(\traj_i, \traj_j; \theta) \geq 0.5, \\
    0\; \text{if}\; P_\surr(\traj_i, \traj_j; \theta) < 0.5.
\end{cases}
\end{equation}

The learning problem is given by
\begin{equation} \label{eq:learn:prob}
    \min\limits_\theta r(\theta) + \frac{1}{\nD} \Sum{\ell=1}{\nD} \cc{L}(p_\ell, P_\surr(\traj_{\cc{I}_\ell}, \traj_{\cc{J}_\ell}; \theta)),
\end{equation}
where $\cc{L}\colon \R \times \R \to \R$ is the cross-entropy loss 
\begin{equation*} \label{eq:learn:func:cross}
        \cc{L}(p, \hat{p}) = -p \log(\hat{p}) - (1 - p) \log(1 - \hat{p})
\end{equation*}
measuring the likelihood of the prediction $\hat{p} \in \R$ matching the target $p \in \{0, 1\}$,
and $r\colon \R^{\nth} \to \R$ is an $\ell_2$-regularization term
$r(\theta) = \rho \| \theta \|^2$,
for a given penalty parameter~$\rho > 0$.\footnote{We note that other regularization terms could be used instead, e.g., $\ell_1$-regularization, or a combination of $\ell_1$ and $\ell_2$ norms.}
The optimal solution $\theta^*$ of~\eqref{eq:learn:prob} provides the parametrization of $\model$ that maximizes the likelihood of the surrogate model $\pi$~\eqref{eq:surrogate} satisfying~\eqref{eq:surr:objective} over the trajectory pairs in the dataset $\Tds$.
Problem~\eqref{eq:learn:prob} can be solved using standard machine learning procedures for different popular choices of the loss function and regularization term, including popular optimization methods such as Adam~\cite{kingma_adam_2014} or L-BFGS-B~\cite{byrd_LBFGS_JSC_1995}; see, e.g., the Python package \texttt{jax-sysid}~\cite{bemporad_TAC_jax-sysid_25}, which uses Adam to warmstart \mbox{L-BFGS-B}.

\subsection{Using the surrogate model for MPC} \label{sec:pref:mpc}

Our choice of taking the surrogate model $\surr$ as in~\eqref{eq:surrogate} and the probability model $P_\surr$ as in~\eqref{eq:sigmoid} differs from the straightforward approach in machine learning for binary classification, cf.~\cite{Dreiseitl_log_reg_2002}.
Indeed, the straightforward approach would be to consider a function $\tilde{\model}\colon \spaceT \times \spaceT \to \R$ parameterized by $\theta$ (typically a NN where $\theta$ collects the weights and biases), and then take~\eqref{eq:sigmoid} as
\begin{equation*}
P_\surr(\traj_i, \traj_j; \theta) = \fracg{1}{1 + \exp(-\tilde{\model}(\traj_i, \traj_j; \theta))}.
\end{equation*}
The surrogate model in this case would be given by~\eqref{eq:surrogate:v2}.

Instead, the idea is to consider a function $\model\colon \spaceT \to \R$ and then take its difference $\model(\traj_i; \theta) - \model(\traj_j; \theta)$ in the exponent term of~\eqref{eq:sigmoid}.
In doing so, we obtain a function $\model(\traj; \theta)$ that can be used to predict the preference between two trajectories as the one given by the trajectory resulting in the smaller value of $\model$, cf.~\eqref{eq:surrogate}.
Therefore, once $\theta^*$ has been obtained by solving the training problem~\eqref{eq:learn:prob}, we obtain a function $\model(\traj; \theta^*)$ such that (the possibly non-unique) $\traj^* \doteq \arg\min_\traj \model(\traj; \theta^*)$ satisfies 
${\surr(\traj^*, \traj; \theta^*) = 1}, \; \forall {\traj \in \spaceT}$.
That is, $\traj^*$ would be preferred (or deemed equal) to any other trajectory $\traj \in \spaceT$ according to our learned surrogate model $\surr$.
The parameter vector $\theta^*$ has been learned so as to increase the likelihood of objective~\eqref{eq:surr:objective} being satisfied for the trajectory pairs in the dataset $\Tds$.
Therefore, assuming that the dataset $\Tds$ is rich enough to capture the underlying preference function $\prefFunc$ and assuming that~\eqref{eq:learn:prob} is successfully solved, the hope is for $\traj^*$ to be preferred by $\prefFunc$ over any other trajectory (not necessarily included in the dataset), that is, for $\prefFunc(\traj^*, \traj) = 1$, $\forall \traj \in \spaceT$.

We can exploit this feature to pose an MPC controller~\cite{Rawlings_MPC_2017, Camacho_S_2013} whose objective function is the learned function $\model(\traj; \theta^*)$:%
\begin{subequations} \label{eq:genMPC}
\begin{align}
    \min\limits_{\traj} \;& \model(\traj; \theta^*) \label{eq:genMPC:cost} \\
    {\rm s.t.} \;& x_0 = x(t), \label{eq:genMPC:init} \\
    &x_{k+1} = f(x_k, u_k), \; k\in\Ni{0}{N-1}, \label{eq:genMPC:dynamics} \\
    & (x_k, u_k) \in \cc{Z}, \; k\in\Ni{0}{N-1}, \label{eq:genMPC:constr} \\
    & x_N \in \cc{X}_N, \label{eq:genMPC:term}
\end{align}
\end{subequations}
where~\eqref{eq:genMPC:init} imposes the initial state constraint, \eqref{eq:genMPC:dynamics} imposes the system dynamics~\eqref{eq:sys} on the predicted states and inputs $(x_k, u_k)$, \eqref{eq:genMPC:constr} impose the state-input constraints on the given set $\cc{Z} \subseteq \R^\nx \times \R^\nu$, and~\eqref{eq:genMPC:term} imposes a terminal constraint on a given terminal set $\cc{X}_N \subseteq \R^\nx$.
Let $\traj^*(t)$ be the optimal solution of~\eqref{eq:genMPC} at time $t$.
The control law of the MPC controller is $u(t) = u_0^*(t)$, where $u_0^*(t)$ corresponds to the first control action in the optimal trajectory $\traj^*(t)$.

By posing the minimization of $\model(\traj; \theta^*)$ as an MPC problem, we have that $\traj^*(t)$ minimizes the objective $\model(\traj; \theta^*)$ for all trajectories satisfying the constraints~\eqref{eq:genMPC:init}-\eqref{eq:genMPC:term}, i.e., for all trajectories satisfying the system dynamics and constraints.
Therefore, we have that
\begin{equation*}
\model(\traj^*(t); \theta^*) \leq \model(\traj; \theta^*),\, \forall \traj \in \spaceT \colon \traj~\text{satisfies}~\eqref{eq:genMPC:init}-\eqref{eq:genMPC:term}.
\end{equation*}
The results is that, according to the learned model $\surr(\traj; \theta^*)$, $\traj^*(t)$ is preferred by $\prefFunc$ over any $\traj \in \spaceT$ satisfying~\eqref{eq:genMPC:init}-\eqref{eq:genMPC:term}.

\begin{remark} \label{rem:stability:MPC}
In this paper, we allow $\model$ to be any function such that problem~\eqref{eq:learn:prob} can be solved using standard machine learning tools.
This includes non-smooth functions $\model$ such as feedforward NN with ReLU activation functions.
In this general setting, the MPC controller~\eqref{eq:genMPC} does not guarantee stability of the closed-loop system.
Additional consideration should be taken to ensure its stability for the choice of $\model$.
\end{remark}

\begin{remark} \label{rem:convexity:MPC}
Problem~\eqref{eq:genMPC} can be particularized to linear MPC, where we note that using $\model$ as the objective function is not restrictive, since it can be taken as a convex function.
Indeed, $\model$ can be taken as a quadratic function, cf.~Section~\ref{sec:results}, or, more generally, as an input-convex neural network~\cite{Amos_in_convex_NN_2017}.
\end{remark}

\begin{remark}
In~Section~\ref{sec:pref:setup} we define $\traj$ as a collection of state, input, and output trajectories.
However, $\model$ does not need to be a function of all three.
That is, $\model$ could be a function of state and inputs $\model(\tX, \tU; \theta)$, inputs and outputs $\model(\tY, \tU; \theta)$, or only consider the initial state $\model(x_0, \tU; \theta)$.
Furthermore, $\traj$ could contain reference information $(x_r, u_r) \in \R^\nx \times \R^\nu$, leading to functions such as $\model(\tX, \tU, x_r, u_r; \theta)$ and an  MPC formulation~\eqref{eq:genMPC} parameterized by $(x_r, u_r)$.
\end{remark}

\section{Numerical results} \label{sec:results}

This section presents numerical results highlighting the proposed preference-based learning approach for MPC objective functions.
We present two case studies: one where the preference function is based on evaluating a quadratic function, and one where we consider a more complex preference function.
Both case studies consider the system of oscillating masses described in~\cite[\S IV.A]{Krupa_TCST_2024}, which is formed by three objects connected by springs.
The state $x = (p_1, p_2, p_3, v_1, v_2, v_3)$ is given by the position $p_i$ and velocity $v_i$ of each object, and the input $u = (F_1, F_2)$ by the forces $F_i$ applied to the two outermost objects.
We take the output as $y = (p_1, p_2, p_3)$.
The dynamics of this system are given by a linear state-space model
\begin{subequations} \label{eq:lin_sys}
\begin{align}
    x(t+1) &= A x(t) + B u(t), \label{eq:lin_sys:x} \\
    y(t) &= C x(t). \label{eq:lin_sys:y}
\end{align}    
\end{subequations}
The control objective is to steer the system to $(x, u) = (0, 0)$ while satisfying the input constraints $|u_i| \leq 1$, $i \in \Ni{1}{2}$.

To solve problem~\eqref{eq:learn:prob}, we start by some initial guess of the parameters $\theta$.
We first perform $200$ iterations of the Adam solver from the \texttt{optax} Python package (version \texttt{0.2.4}), taking the learning rate as $0.01$ and decay rates as $\beta_1 = 0.9$, $\beta_2 = 0.999$.
We then run a maximum of $1000$ iterations of the L-BFGS-B solver (with a history size of $10$) from the \texttt{jaxopt} Python package (version \texttt{0.8.3}), starting from the solution returned by Adam.
We take $\rho = 10^{-6}$ for the $\ell_2$-regularization term of~\eqref{eq:learn:prob}.
Tests are performed using Python \texttt{3.11.10} on a Macbook Air with an M3 microprocessor.

\subsection{Learning a quadratic-based preference function} \label{sec:results:quad}

We first evaluate the proposed approach by considering a scenario where the preference function is given by
\begin{equation} \label{eq:results:quad:prefFunc}
\prefFunc_\phi(\traj_j, \traj_j) =
\begin{cases}
    1\; \text{if}\; \phi_N(\traj_i; Q, R) \leq \phi_N(\traj_j; Q, R), \\
    0\; \text{if}\; \phi_N(\traj_i; Q, R) > \phi_N(\traj_j; Q, R),
\end{cases}
\end{equation}
with
\begin{equation} \label{eq:quad:func}
    \phi_N(\traj; Q, R) = \Sum{k=0}{N-1} \| x_k \|^2_Q + \| u_k \|^2_R,
\end{equation}
for unknown positive definite matrices $Q \in \R^{\nx \times \nx}$ and $R \in \R^{\nu \times \nu}$.
That is, $\prefFunc_\phi$ prefers the trajectory with the smallest quadratic cost for matrices $Q$, $R$, and horizon $N$, which we take in our simulations as $Q = \texttt{diag}(40, 40, 40, 5, 5, 5)$, $R = \texttt{diag}(0.2, 0.2)$, and $N = 10$.

We build a dataset $\Tds$ by first generating $n_t = 50$ closed-loop trajectories of~\eqref{eq:lin_sys}, each starting from a random $x(0)$ and controlled using the Linear Quadratic Regulator (LQR) for random matrices $\hat{Q} \in \R^{\nx \times \nx}$ and $\hat{R} \in \R^{\nu \times \nu}$ penalizing states and inputs, respectively.
For $x(0)$, we take the position and velocity of each object from uniform distributions in the intervals $[-0.3, 0.3]$ and $[-0.05, 0.05]$, respectively.
For $\hat{Q}$ and $\hat{R}$ we generate predominantly-diagonal positive definite matrices, with diagonal elements in the range $[0.1, 10]$.
Datasets of paired trajectories are obtained by randomly selecting pairs from the $n_t$ trajectories and then evaluating the preference function on each pair.

{\renewcommand{\arraystretch}{1.0}%
    \begin{table}[t]
    \setlength{\tabcolsep}{4.0pt}
    \centering
    \begin{threeparttable}
\caption{Training and performance results for $\prefFunc_\phi$.}
    \label{tab:results:quad}
    \begin{tabular}{lcccccc}
        \toprule
        MPC cost & Train & Test & Time [s] & Avrg. $\phi_{30}$ & Max. $\phi_{30}$ & Min. $\phi_{30}$ \\
        \cmidrule(lr){1-1} \cmidrule(lr){2-3} \cmidrule(lr){4-4} \cmidrule(lr){5-7}
                             From $\prefFunc$ & -- & -- & -- & 1.000 & 4.426 & 0.054 \\
                             Random & -- & 82.6 & -- & 1.833 & 10.532 & 0.106 \\
                             $\model_{20}$ & 100.0 & 92.0 & 1.13 & 1.355 & 5.491 & 0.086 \\
                             $\model_{60}$ & 100.0 & 94.8 & 1.13 & 1.108 & 4.598 & 0.059 \\
                             $\model_{100}$ & 98.0 & 95.0 & 1.16 & 1.230 & 5.192 & 0.059 \\
                             $\model_{400}$ & 100.0 & 99.4 & 1.53 & 1.043 & 4.536 & 0.057 \\
                             $\model_{1000}$ & 100.0 & 99.8 & 1.87 & 1.013 & 4.404 & 0.056 \\
        \bottomrule
    \end{tabular}
\begin{tablenotes}[flushleft]
\scriptsize
\item ``Train", and ``Test" are the accuracy on training and test datasets, in \%. ``Time" is the average computation time of solving problem~\eqref{eq:learn:prob}. The average, maximum and minimum performance, with performance measured by $\phi_{30}$, have all been normalized by the average performance obtained with the MPC using the $Q$ and $R$ from $\prefFunc_\phi$.
\end{tablenotes}
\end{threeparttable}
\end{table}}

\begin{figure*}[t]
    \centering
    \begin{subfigure}[ht]{0.32\textwidth}
        \includegraphics[width=\linewidth]{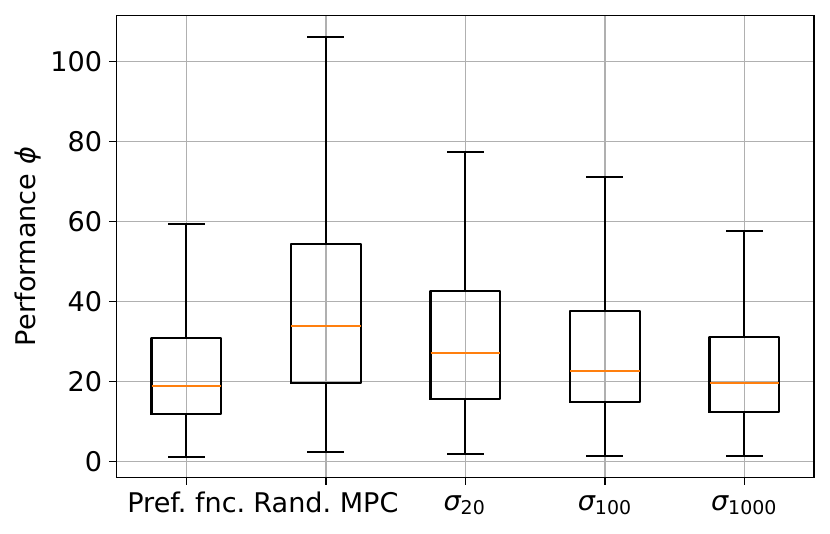}
        \caption{Closed-loop performance $\phi_{30}$.}
        \label{fig:results:quad:perf}
    \end{subfigure}%
    \hfill
    \begin{subfigure}[ht]{0.32\textwidth}
        \includegraphics[width=\linewidth]{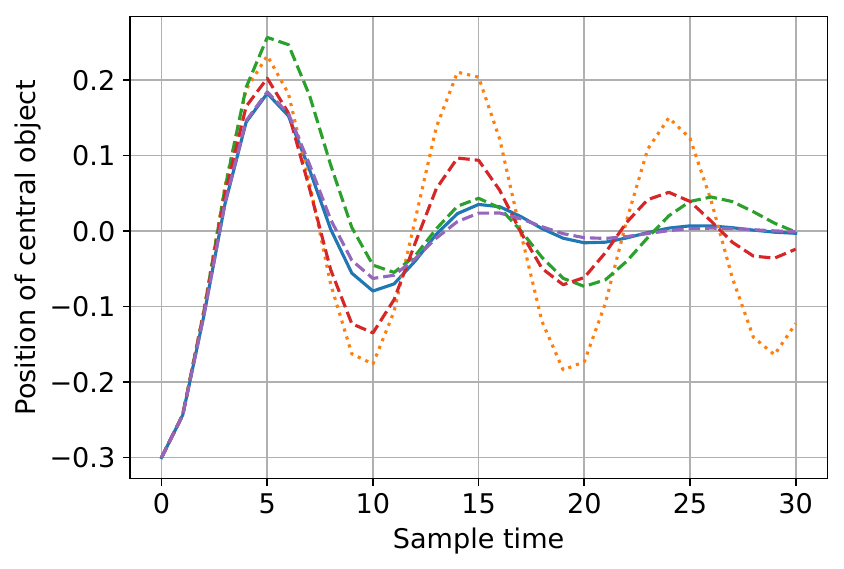}
        \caption{Closed-loop state trajectory.}
        \label{fig:results:quad:state}
    \end{subfigure}%
    \hfill
    \begin{subfigure}[ht]{0.32\textwidth}
        \includegraphics[width=1\linewidth]{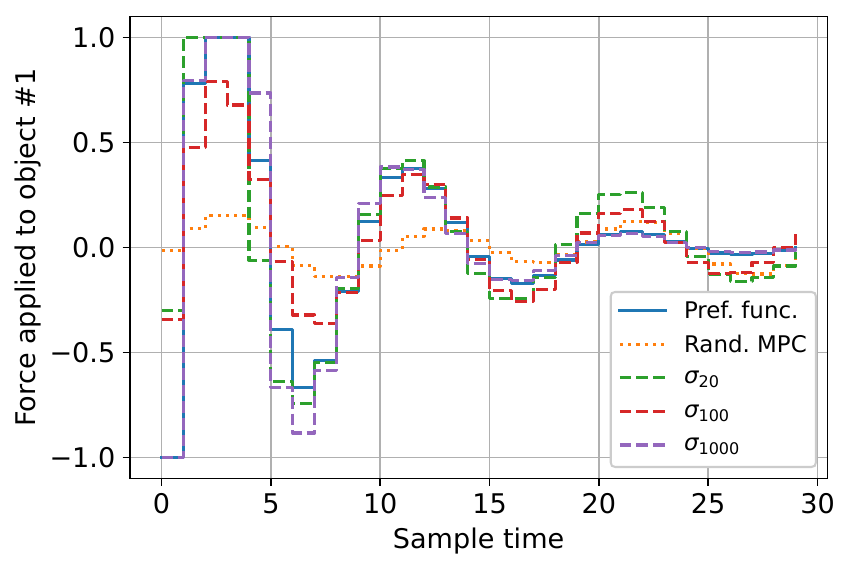}
        \caption{Closed-loop input trajecory.}
        \label{fig:results:quad:input}
    \end{subfigure}%
    \hfill
    \caption{Results for quadratic-based preference function $\prefFunc_\phi$ and quadratic $\model$.}
    \label{fig:results:quad}
\end{figure*}

We take $\model$ as a quadratic function~\eqref{eq:quad:func}:
\begin{equation} \label{eq:model:quad}
    \model(\traj; \theta) = \phi_N(\traj; Q_\theta, R_\theta),
\end{equation}
where $Q_\theta \in \R^{\nx \times \nx}$ and $R_\theta \in \R^{\nu \times \nu}$ are encoded by~$\theta$ as follows.
Let $n_Q = \nx + \frac{\nx (\nx -1)}{2}$ and $n_R = \nu + \frac{\nu (\nu - 1)}{2}$.
Then $\theta \in  \R^{n_Q + n_R}$ contains the non-zero elements of the Cholesky decomposition of matrices $Q_\theta$ and $R_\theta$.
This encoding allows us to impose positive definiteness on matrices $Q_\theta$ and $R_\theta$ by enforcing $\theta_{(i)} \geq \epsilon$, with $\epsilon > 0$, for all components $i$ of $\theta$ corresponding to an element in the diagonal of $Q_\theta$ or $R_\theta$. We take $\epsilon = 0.01$
This lower-bound constraint on $\theta$ is imposed when solving problem~\eqref{eq:learn:prob}.
Additionally, we impose the  first element in the diagonal of $Q_\theta$ to be greater or equal to $1$, which is formulated as a lower bound on the corresponding element of $\theta$ in~\eqref{eq:learn:prob}.\footnote{This additional constraint is added to limit the degree of freedom in scaling matrices $Q_\theta$ and $R_\theta$, while also improving numerical conditioning by avoiding small values in $\theta^*$.}

By taking $\model$ as in~\eqref{eq:model:quad}, we guarantee the existence of matrices $Q_\theta$ and $R_\theta$ such that our model perfectly captures the preference function~\eqref{eq:results:quad:prefFunc}.
Indeed,~\eqref{eq:surr:objective} is always satisfied if $Q_\theta = Q$ and $R_\theta = R$.
However, we assume that matrices $Q$ and $R$ are unknown by the learning process to simulate a real scenario where the underlying preference function is unknown.
Therefore, we randomly initialize $Q_\theta$ and $R_\theta$ using the same approach taken to generate matrices $\hat{Q}$ and $\hat{R}$ for the dataset $\Tds$.
The $\theta$ corresponding to these random $Q_\theta$ and $R_\theta$ is used to initialize the Adam solver.

We consider training datasets of different sizes and a testing dataset with $500$ paired trajectories, which we use to evaluate the accuracy of the trained surrogate models.
Model accuracy is measured as
\begin{equation*}
    \frac{1}{\nD} \Sum{\ell = 1}{\nD} \left[ \surr(\traj_{\cc{I}_\ell}, \traj_{\cc{J}_\ell}; \theta^*) = \prefFunc(\traj_{\cc{I}_\ell}, \traj_{\cc{J}_\ell}) \right],
\end{equation*}
where $\nD$ is the number of trajectory pairings $(\traj_{\cc{I}_\ell}, \traj_{\cc{J}_\ell})$ in the test dataset.
For each dataset size $\nD$, we train $20$ functions $\model_\nD$, each one starting from a different random initialization of $\theta$.
We keep the $\model_\nD$ with the highest accuracy on the test dataset.
We then use the functions $\sigma_\nD$ as objective functions of the following particularization of the MPC controller~\eqref{eq:genMPC}:
\begin{subequations} \label{eq:stanMPC}
\begin{align}
    \min\limits_{\traj} \;& \phi_N(\traj; \hat{Q}, \hat{R}) \label{eq:stanMPC:cost} \\
    {\rm s.t.} \;& x_0 = x(t), \label{eq:stanMPC:init} \\
    &x_{k+1} = A x_k + B u_k, \; k\in\Ni{0}{N-1}, \label{eq:stanMPC:dynamics} \\
    & \underline{u} \leq u_k \leq \overline{u} \in, \; k\in\Ni{0}{N-1}. \label{eq:stanMPC:u_bound}
\end{align}
\end{subequations}
We perform $200$ closed-loop simulations with system~\eqref{eq:lin_sys}, each one starting from a random $x(0)$ computed as for the generation of the datasets.
Simulations have a length of $30$ sample times.
Additionally, for the purpose of comparison, we take an MPC~\eqref{eq:stanMPC} with $\hat{Q} = Q$ and $\hat{R} = R$, and, for each simulation, an MPC~\eqref{eq:stanMPC} with random matrices $\hat{Q}$ and~$\hat{R}$.

Table~\ref{tab:results:quad} shows the training and test results.
We report the accuracy of the models on training and test datasets, as well as the average computation time of solving~\eqref{eq:learn:prob}.\footnote{This includes the execution of Adam, L-BFGS-B and overhead.}
We also report the average, maximum and minimum performance of the closed-loop trajectories for each MPC controller, with performance measured as $\phi_{30}(\traj; Q, R)$.
That is, performance is measured using matrices $Q$ and $R$ from the preference function~\eqref{eq:results:quad:prefFunc}. 
Therefore, a smaller value of the performance index corresponds to a closed-loop trajectory that would be preferred by $\prefFunc_\phi$~\eqref{eq:results:quad:prefFunc}.
We normalize all performance values by the average performance obtained using the MPC controller with $Q$ and $R$ from the preference function $\prefFunc_\phi$~\eqref{eq:results:quad:prefFunc}.
Fig.~\ref{fig:results:quad} shows performance results of the MPC controllers as well as an example comparing the closed-loop trajectories obtained by different MPC controller for a random initial state $x(0)$ with position of the central object set to $-0.3$.

The results show that as the size $\nD$ of the training dataset increases, the learned functions $\model_\nD$ tend to provide a higher accuracy on the $500$ trajectory pairs within the test dataset.
That is, we obtain matrices $Q_\theta$ and $R_\theta$ that capture the preference function $\prefFunc_\phi$.
We note that matrices $Q_\theta$ and $R_\theta$ obtained in these tests do not necessarily resemble matrices $Q$ and $R$.
In particular, $Q$ and $R$ are diagonal matrices, whereas $Q_\theta$ and $R_\theta$ typically have significant non-diagonal values.
Nevertheless, the MPC controllers using $\model_\nD$ provide good closed-loop performance.
Indeed, results indicate that the use of functions $\model_\nD$ as objective function of the MPC~\eqref{eq:stanMPC} provide a closed-loop behavior and performance that resemble the ones obtained by using the $Q$ and $R$ from $\prefFunc_\phi$~\eqref{eq:results:quad:prefFunc}, when using a sufficiently large training dataset.
Note that for $\model_{1000}$, the average performance only increases by $1.3\%$.
We also observe that the use of functions $\model_\nD$ obtained from very small datasets provide better results than using MPC controllers with random matrices $\hat{Q}$ and $\hat{R}$ (generated with the same procedure used to populate the dataset).

\subsection{Learning a complex preference function} \label{sec:results:complex}

{\renewcommand{\arraystretch}{1.0}%
    \begin{table*}[t]
    \setlength{\tabcolsep}{4.0pt}
    \centering
\caption{Training and performance results for complex preference function $\prefFunc_\settime$.}
    \label{tab:results:complex}
    \begin{tabular}{lrrccccc}
        \toprule
        \multicolumn{1}{l}{}& \multicolumn{3}{c}{Training and test results after solving~\eqref{eq:learn:prob}} & \multicolumn{2}{c}{Results for $\prefFunc_\settime$ with $\prefFunc_u$} & \multicolumn{2}{c}{Results for $\prefFunc_\settime$ without $\prefFunc_u$} \\
        \cmidrule(lr){2-4}\cmidrule(lr){5-6}\cmidrule(lr){7-8}
        MPC cost & Acc. train (\%) & Acc. test (\%) & Avrg. comp. time [s] & Med./Max. $\settime$ & Avrg. $\max \|u(t)\|_\infty$ & Med./Max. $\settime$  & Avrg. $\max \|u(t)\|_\infty$ \\
        \cmidrule(lr){1-1} \cmidrule(lr){2-8}
                             Random & -- & 83.17 & -- & 10 / 21 & 0.52 & 10 / 21 & 0.52 \\
                             $\model_{20}$ & 100.0 & 90.60 & 1.35 & 16 / $>$30 & 0.51 & 10 / 16 & 1.53 \\
                             $\model_{100}$ & 100.0 & 93.60 & 1.43 & 9 / 14 & 3.21 & 7 / 9 & 3.06 \\
                             $\model_{400}$ & 95.25 & 94.80 & 1.91 & 9 / 17 & 2.38 & 5 / 9 & 5.48 \\
                             $\model_{1000}$ & 95.20 & 94.20 & 2.75 & 8 / 11 & 2.38 & 5 / 6 & 6.38 \\
        \bottomrule
    \end{tabular}
\end{table*}}

\begin{figure*}[t]
    \centering
    \begin{subfigure}[ht]{0.32\textwidth}
        \includegraphics[width=\linewidth]{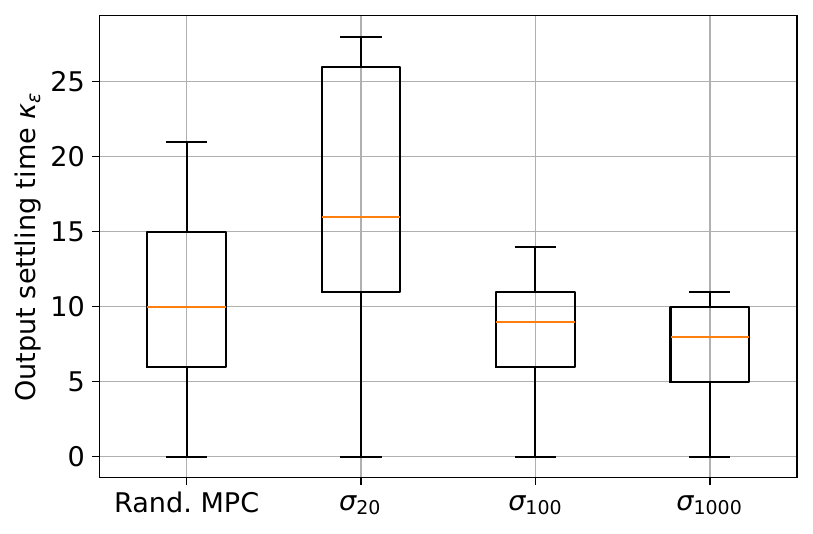}
        \caption{Closed-loop values of $\settime$.}
        \label{fig:results:complex:box}
    \end{subfigure}%
    \hfill
    \begin{subfigure}[ht]{0.32\textwidth}
        \includegraphics[width=\linewidth]{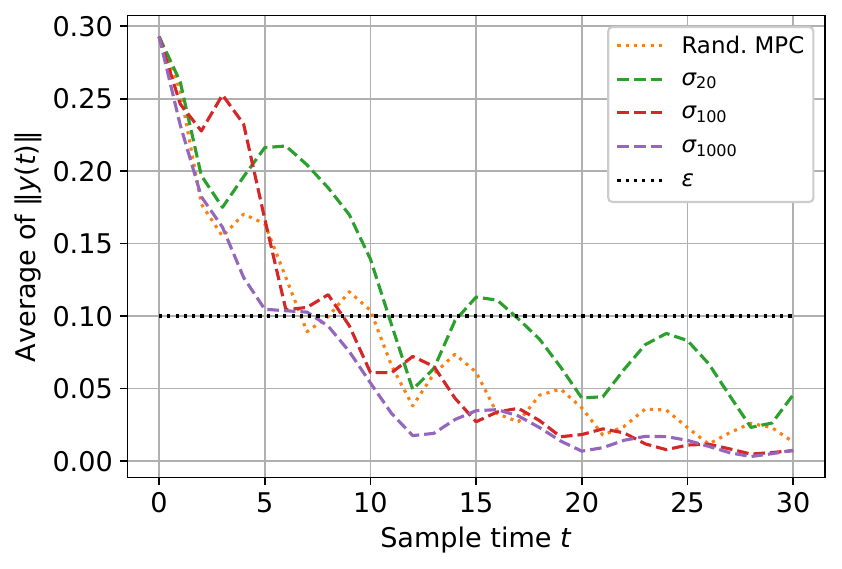}
        \caption{Avrg. of $\|y(t)\|$ with $\max \|u(t)\|_\infty$.}
        \label{fig:results:complex:max_u}
    \end{subfigure}%
    \hfill
    \begin{subfigure}[ht]{0.32\textwidth}
        \includegraphics[width=1\linewidth]{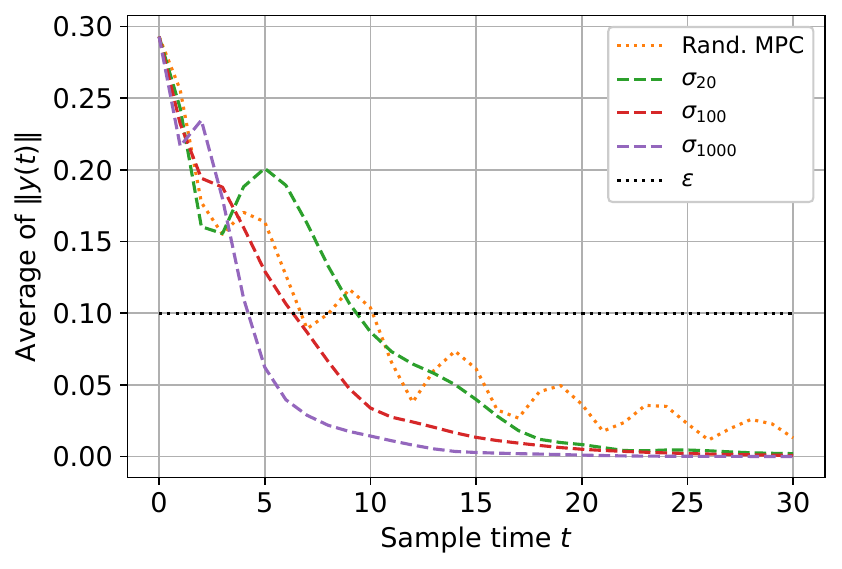}
        \caption{Avrg. of $\|y(t)\|$ without $\max \|u(t)\|_\infty$.}
        \label{fig:results:complex:non_max_u}
    \end{subfigure}%
    \hfill
    \caption{Results for preference function $\prefFunc_\settime$ and quadratic $\model$.}
    \label{fig:results:complex}
\end{figure*}

We now consider a preference function that selects the trajectory with the smallest $2$-norm output settling time:
\begin{equation*}
    \settime(\traj) \doteq \min\limits_{k \geq 0} k \colon k \in \N, \| y_k \| \leq \varepsilon, \| y_\ell \| \not> \varepsilon\; \forall \ell > k,
\end{equation*}
where we take $\varepsilon = 0.1$.
That is, we consider
\begin{equation*}
\prefFunc_\settime(\traj_j, \traj_j) =
\begin{cases}
    1\; \text{if}\; \settime(\traj_i) < \settime(\traj_j), \\
    0\; \text{if}\; \settime(\traj_i) > \settime(\traj_j), \\
    \prefFunc_u(\traj_j, \traj_i) \; \text{otherwise},
\end{cases}
\end{equation*}
with
\begin{equation*}
\prefFunc_u(\traj_j, \traj_j) =
\begin{cases}
    1\; \text{if}\; \max \|u_i(t)\|_\infty \leq \max \|u_j(t)\|_\infty, \\
    0\; \text{otherwise},
\end{cases}
\end{equation*}
where $u_i(t)$ and $u_j(t)$ are the input trajectories associated with $\traj_i$ and $\traj_j$, respectively.
The combination of $\prefFunc_\settime$ with $\prefFunc_u$ is designed to predominantly choose trajectories with small output settling times $\settime$ while encouraging the use of small maximum values of the control action.

As in Section~\ref{sec:results:quad}, we obtain a dataset of paired trajectories from closed-loop simulations with random LQR controllers.
In this case, we take $N = 15$,
and $\hat{Q}$ and $\hat{R}$ as diagonal matrices with non-zero elements taken from uniform distributions on the following intervals: $[5, 20]$ for elements of $Q$ corresponding to object positions, and $[0.1, 1.0]$ for the rest.
This change provides datasets with a rich variety of values of $\settime$ that typically satisfy $\settime(\traj) \leq N$ for all $\traj$ in the dataset.
The median value of $\settime$ for the generated trajectories is $6$.
As in Section~\ref{sec:results:quad}, we take $\model$ as a quadratic function~\eqref{eq:model:quad} and perform $200$ closed-loop simulations with the MPC controllers~\eqref{eq:stanMPC} taking random initial states $x(0)$.
In this case, we remove the input constraints~\eqref{eq:stanMPC:u_bound} so as to provide the controller with freedom to achieve smaller output settling times $\settime$.

Table~\ref{tab:results:complex} shows the training results, the median and maximum values of $\settime$ for the closed-loop tests, and the average value of $\max \|u(t)\|_\infty$.
We also report the results on $\settime$ and $\max \|u(t)\|_\infty$ for the same tests but removing $\prefFunc_u$ from $\prefFunc_\settime$.\footnote{We remove from the dataset all trajectory pairs with the same $\settime$.}
Fig.~\ref{fig:results:complex} shows the results for the output settling time and the average values of $\| y(t) \|$ for each sample time $t$ of the simulations.
It also shows the average values of $\| y(t) \|$ obtained when removing $\prefFunc_u$ from $\prefFunc_\settime$.
The results show that we learn objective functions that tend to provide smaller output settling times $\settime$ as the size of the dataset increases.
Additionally, we observe that removing $\prefFunc_u$ from $\prefFunc_\settime$ results in smaller values of $\settime$ at the expense of an increase in the control actions, as expected.

\section{Conclusions} \label{sec:conclusions}

We have presented a preference-based learning approach for learning the objective function of an MPC controller from human preferences expressed over trajectory pairs.
We learn a surrogate function of the underlying human preference function by posing a binary classification problem, which can be solved using standard machine learning tools.
The problem is posed so that the fitted model can be directly used as the objective function of an MPC controller so as to provide a closed-loop behavior that aligns with the human preferences.
Case studies show good results when using $50$ closed-loop trajectories and a few hundred evaluations of the preference function, indicating that the MPC can be tuned using a relatively small amount of machine and human~time.

\bibliographystyle{IEEEtran}
\bibliography{IEEEabrv,bib_prefLearnMPC}

\end{document}